# Activated carbon is an electron-conducting amphoteric ion adsorbent

P.M. Biesheuvel

[1]*Wetsus, European Centre of Excellence for Sustainable Water Technology, Leeuwarden, The Netherlands.*

[2]*Laboratory of Physical Chemistry and Soft Matter, Wageningen University, The Netherlands.*

E-mail: maarten.biesheuvel@wetsus.nl

**Abstract**

Electrodes composed of activated carbon (AC) particles can desalinate water by ion electrosorption. To describe ion electrosorption mathematically, accurate models are required for the structure of the electrical double layers (EDLs) that form within electrically charged AC micropores. To account for salt adsorption also in uncharged ACs, an "attraction term" was introduced in modified Donnan models for the EDL structure in ACs. Here it will be shown how instead of using an attraction term, chemical information of the surface structure of the carbon-water interface in ACs can be used to construct an alternative EDL model for ACs. This EDL model assumes that ACs contain both acidic groups, for instance due to carboxylic functionalities, and basic groups, due to the adsorption of protons to the carbon basal planes. As will be shown, this "amphoteric Donnan" model accurately describes various data sets for ion electrosorption in ACs, for solutions of NaCl, of $CaCl_2$, and mixtures thereof, as function of the external salt concentration and of the cell voltage between two AC electrodes in a process called capacitive deionization (CDI). The amphoteric Donnan model can be used to relate the conditions of activation treatment and electrode preparation (which both influence the carboxylic acid content) to the EDL-structure and desalination performance of ACs.

**Introduction**

Water can be desalinated with porous carbon electrodes by the phenomenon of ion electrosorption.[1-5] Electrosorption is used in a process called capacitive deionization (CDI) where upon transferring electronic charge through an external circuit from one porous electrode to the other, electrical double layers (EDLs) are formed within the micropores of the carbon electrodes. In these micropores, electronic charge (in the carbon) is locally charge-compensated by ionic charge in the water that fills the pores. This ionic charge is due to a difference in concentration between counterions and coions, see Fig. 1.[6-8] During cell charging, more counterions are adsorbed in the EDLs than coions are expelled, and the water flowing through a CDI device (consisting of two porous electrodes) is desalinated. The cell voltage $V_{cell}$ equals the anode potential minus the cathode potential, and switches during CDI cell operation between the charging voltage (for instance, $V_{ch}$=1.2 V), and the discharge voltage (typically $V_{disch}$=0 V). In this cyclical operational scheme, freshwater is produced for some time (during the charging step), after which for some time water of a higher salinity is produced (discharge step).

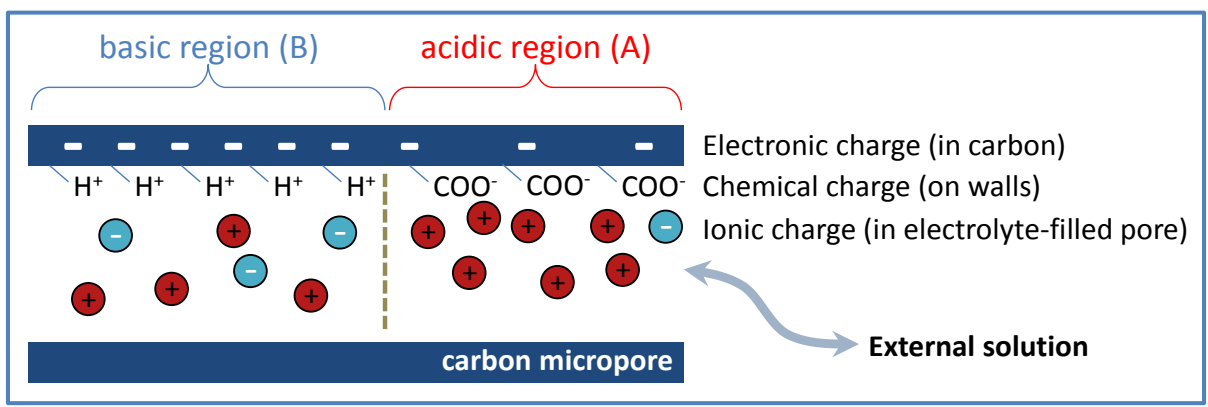

Fig. 1. Sketch of EDL structure in a micropore in activated carbon according to the "amphoteric Donnan" model. On the scale of nanometers or less, regions influenced by adsorbed protons (basic, "B") are distinguished from regions influenced by carboxylic groups (acid, "A"). The electronic charge in the carbon can differ between the two regions, as well as the ionic composition in the aqueous phase. Both types of regions are in equilibrium with the external solution outside the micropore.

To semi-analytically describe the EDL-structure in AC micropores, various models have been proposed over the years, either based on the Gouy-Chapman-Stern (GCS) framework,[7,9,10] or using the Donnan model.[3,5,6,11-13] The Donnan model is a useful simplification of GCS-theory as it assumes full overlap of the diffuse layers and thus the electrical potential no longer depends on the exact position in the pore. For micropores (size < 2 nm) in water of a salinity of 100 mM or less (Debye length 1 nm or more), this seems a realistic simplification. To make the Donnan model work, a Stern capacity is added, describing a voltage drop between the locations of electronic and ionic charge, and in addition a "chemical attraction term" is included, which is an attractive force for ions to go into the micropores. This attraction term was first taken as a constant,[11] but in later work assumed to be inversely proportional to the total ion concentration in the pore.[6,13] By including this term, various data sets of salt adsorption vs. cell voltage and external salinity could be very well described by this modified Donnan (mD) model. In addition to describing these data, the mD model including the attraction term correctly predicted that uncharged carbon also absorbs salt, though the fit to available data was not very good (Fig. 6 in ref. [6]).

Recently we developed a novel Donnan model that describes and predicts various novel modes of CDI-operation, such as inverted-CDI and enhanced-CDI, as well as explaining the "inversion peak" often found in a CDI cycle at the start of the discharge step.[14-16] In this theory, the key modification to existing Donnan models was the inclusion of fixed, immobile, chemical charge in one or both of the electrodes, while the attraction term was dropped.[16] This extended EDL-model proved to be a breakthrough for the mathematical description of ion electrosorption because it was able to explain the abovementioned counterintuitive phenomena. However, other experimental observations related to ion electrosorption could not yet be described, such as the nonzero salt adsorption capacity of uncharged carbon powders.

From studies of the chemistry of carbon,[2,17,18] it is known that not only acidic groups of various kinds are present in activated carbon micropores (most notably carboxylic acid, but also phenolic groups can respond as an acid), whose concentration depends on the thermal or chemical activation process, and on the solvent used in electrode manufacturing, but the carbon basal planes themselves also interact with the water and respond as a base.[17,18] This is due to their propensity to absorb protons, resulting in a positive chemical charge at the surface. Thus, AC micropores contain both positive and negative chemical charge,[19,20] and therefore activated carbons fall in the class of amphoteric materials.

In the present note, we show how published data on CDI for monovalent and asymmetric salts can be well described by the "amphoteric Donnan" (amph-D) model which considers two types of fixed chemical charge in the carbon micropores. The amph-D model does not use the previously introduced attraction term, but instead is based on realistic values of the negative and positive chemical charge, whose values can be measured independently by titration. Data for uncharged carbons (Fig. 6 in ref. [6]) are now very well described without additional parameters.

**Theory**

The "amphoteric-Donnan" (amph-D) model extends the Donnan model for carbon micropores[3,5,6,13-12] by including fixed chemical charge.[16] However, whereas ref. [16] only considered *one* type of fixed chemical charge in the pores, here we will assume the presence of *two* types of chemical charge -- one of positive sign (basic groups, abbreviated as "B") and one of negative sign (acidic groups, "A"). Each electrode, both the anode and cathode, generally will have both of these groups in their electrodes (we do not for instance assume A-groups only in the anode and B-groups in the cathode). The general theory allows for the chemical charge densities to have any value and sign. There are also no restrictions on the distribution of pore volume over which part is "under the influence of" the A-groups, and which part of the volume is near the B-groups, see Fig. 1.

This brings us to the following important point about the amph-D model. For this model to work, we must make the assumption that the electrostatic environment around A-groups is separate from that around B-groups, and is not "smeared-out" with the two types of chemical charge (two types of regions) charge-compensating one another with one effective chemical charge and potential remaining. Similar models based on separated ionic environments on the (sub)-nm scale are also proposed for ion adsorption in membranes for nanofiltration[21] and electrodialysis.[22] Thus, instead of assuming a smeared-out potential, we consider that in a carbon micropore an ion will either find itself in an "A-region" or a "B-region." These regions, though possibly just fractions of a nm apart, must be assumed in the theory to create separated electrostatic environments with different electrostatic potentials. Note how in various limits the amph-D model simplifies to existing models: in case we assume only one type of micro-region to exist, then the model of ref. [16] results, while setting all values of the fixed chemical charge to zero will reduce the amph-D model to the most simple Donnan approach without the chemical attraction term.[12]

In general, in each micro-region in each electrode, the ionic charge, $\sigma_{ionic}$, together with the chemical charge, and the charge in the electronic phase, i.e., in the carbon matrix, $\sigma_{elec}$, charge-compensate one another, see Fig. 1. Thus,[23]

$$\sigma_{ionic,j} + \sigma_{chem,j} + \sigma_{elec,j} = 0 \tag{1}$$

where subscript "j" refers to either an A- or B-region.

To describe the EDL structure in each micro-region we use the Donnan model which considers that in carbon micropores (< 2 nm), the EDLs forming along pore surfaces strongly overlap. Thus, a single electrostatic potential and concentration in each micro-region is assumed, with the ion concentration in the pore related to concentrations outside the pore by Boltzmann's law. The potential difference between inside and outside the micropore is the Donnan potential, $\Delta\phi_D$, which will be different for the A- and B-regions. An additional potential drop, $\Delta\phi_S$, again different between A- and B-regions, is assumed to depend linearly on the electronic charge, and may be due to Stern layer effects or a quantum capacity (space charge layer) within the carbon, see ref. [16] for further references. Note that dimensionless potentials, $\phi$, can be multiplied by the thermal voltage, $V_T$ ($=RT/F$) to arrive at voltages, $V$ or $E$, with units of Volts.

For a 1:1 salt, at equilibrium, potentials $\Delta\phi_D$ and $\Delta\phi_S$ relate to the ionic and electronic charge density according to

$$\Delta\phi_{D,j} = -\text{arcsinh}\left(\sigma_{\text{ionic},j}/2c_\infty\right) \quad , \quad \Delta\phi_{S,j} = \sigma_{\text{elec},j} \cdot F/C_S \cdot V_T \qquad (2)$$

where $c_\infty$ is the external salt concentration, i.e., outside the micropores, and $C_s$ is the Stern capacity, while all charge densities, $\sigma_j$, are defined per unit pore volume. In the Donnan model, the total ion concentration in each micro-region, $c_{\text{ions},j}$, is given by[6]

$$c_{\text{ions},j}^2 = \sigma_{\text{ionic},j}^2 + \left(2 \cdot c_\infty\right)^2. \qquad (3)$$

Within one and the same electrode the two micro-regions are coupled together by

$$c_{\text{ions,mi}} = \sum_{j=A,B} \alpha_j c_{\text{ions},j} \quad , \quad \sigma_{\text{elec}} = \sum_{j=A,B} \alpha_j \sigma_{\text{elec},j} \qquad (4)$$

where $c_{\text{ions,mi}}$ is the micropore-averaged total ions concentration, $\alpha_j$ is the fraction of the total micropore volume, $v_{\text{mi}}$ (in mL/g electrode), that is in the respective micro-region, and where $\sigma_{\text{elec}}$ is the pore-averaged electronic charge density in the carbon. These pore-averaged concentrations, $c_{\text{ions,mi}}$ and $\sigma_{\text{elec}}$, can be used in two-electrode models for ion electrosorption and CDI as before.[6,13,12,16] For asymmetric salts such as CaCl$_2$ (where $z_+$=2 and $z_-$=-1) and for mixtures of monovalent and asymmetric salts, Eqs. (2a), (3) and (4a) must be replaced by the Boltzmann equation,

$$c_{i,j} = c_{\infty,i} \cdot \exp\left(-z_i \cdot \Delta\phi_{D,j}\right) \qquad (5)$$

to be evaluated for all ions, i, anions and cations, in both micro-regions, j, together with

$$c_{i,\text{mi}} = \sum_{j=A,B} \alpha_j c_{i,j} \qquad (6)$$

for the micropore-averaged concentration of ion type i.

For a 1:1 salt, and when $v_{\text{mi}}$ is the same in both electrodes, the salt adsorption, commonly expressed in mg salt adsorption per dry mass (in g) of both anode and cathode together, is calculated by comparing the values of $c_{\text{ions,mi}}$ in both electrodes at the end of the charging and discharge steps, summing over both electrodes, and multiplying by ¼·$M_w$·$v_{\text{mi}}$ where $M_w$ is the salt molar mass ($M_w$=58.44 g/mol for NaCl), which results in the salt adsorption capacity ($\Gamma_{\text{salt}}$ or SAC) of the two-electrode cell.[5] For charge $\Sigma_F$ in C/g, we multiply $\sigma_{\text{elec}}$ in either electrode with $v_{\text{mi}}$ and the factor ½. The ratio of SAC over $\Sigma_F$ (corrected by a factor $M_w/F$) is the charge efficiency, $\Lambda$, which is between 0 and 1. For 2:1 salts and mixtures, we sum $c_{i,\text{mi}}$ over both electrodes, compare values after charging and discharge, and multiply by ½·$v_{\text{mi}}$, to obtain the adsorption of ion i by the electrode pair in mol/g.

**Results and Discussion**

To validate the amph-D model, we compare calculation results with experimental data presented in ref. [6] for NaCl, and presented in ref. [12] for $CaCl_2$ and mixtures of NaCl and $CaCl_2$. As an arbitrary choice, in all calculations we assign equal micropore volumes to the acidic and the basic regions, i.e., the parameter $\alpha$ used in Eq. (4) is set to $\alpha=\frac{1}{2}$. Furthermore, except for results presented in Fig. 6, again as an arbitrary choice, we give the acidic and basic chemical charge the same absolute value, i.e., $\sigma_{chem,A}+\sigma_{chem,B}=0$.

Experimental results for NaCl salt adsorption in uncharged carbon powders reported in ref. [6] are reproduced in Fig. 2 and show a broad maximum in adsorption for external salinities, $c_\infty$, around 10-40 mM, beyond which salt adsorption slowly drops. In ref. [6], these data could not be described well by the "improved modified" Donnan model, a model which includes an attraction term $\mu_{att}$ that inversely depends on total micropore ions concentration. Here, we report the fit of the amph-D model to the same data set, and as Fig. 2 shows, excellent agreement is now obtained when we use a value for the chemical charge in the micropores of $\sigma_{chem,A/B}=\pm 0.26$ M. This value of $\sigma_{chem}$ is very realistic, being about three times less than the maximum concentration of chemical groups reported in Fig. 6 of ref. [17] (assuming $v_{mi}=0.5$ mL/g). The value of $\sigma_{chem}=\pm 0.26$ M will be used in all data analysis in the present work. Note that for uncharged carbons, with the assumptions as stated (namely, $\alpha=\frac{1}{2}$ and $\sigma_{chem,A}+\sigma_{chem,B}=0$), Eqs. (1)-(4) can be simplified significantly and we only need to find a numerical solution of

$$\Delta\phi_D + \Delta\phi_S = 0 \quad , \quad 2c_\infty \cdot \sinh(\Delta\phi_D) + \sigma_{chem} - \Delta\phi_S \cdot C_S V_T / F = 0 \quad , \quad c_{salt,exc} = c_\infty \cdot (\cosh(\Delta\phi_D) - 1). \tag{7}$$

For high salinity, the Donnan potential is low, and Eq. (7) simplifies to

$$\Delta\phi_D = \frac{\sigma_{chem}}{2c_\infty + C_S V_T / F} \quad , \quad c_{salt,exc} = \tfrac{1}{2} c_\infty \cdot \Delta\phi_D^2. \tag{8}$$

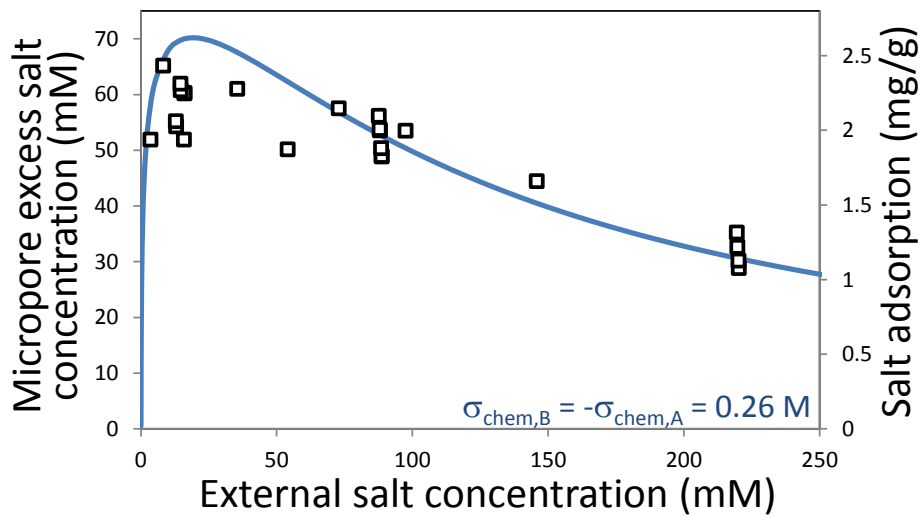

Fig. 2. Excess salt adsorption in activated carbon. Data (squares) from ref. [6] and theory (solid line) based on Eq. (7). Input parameters in Table 1.

Table 1. Parameter settings.

| | | | |
|---|---|---|---|
| Stern capacity | $C_S$ | GF/m$^3$ | 0.145 (from ref. [6]) |
| Fixed chemical charge | $\sigma_{chem,A}$ | M | -0.26 |
| | $\sigma_{chem,B}$ | M | +0.26 |
| Volume ratio micro-region | $\alpha_A=\alpha_B$ | | ½ |
| Micropore volume | $v_{mi}$ | mL/g_AC | 0.64 (data Fig. 2, ref. [6]) |
| | $v_{mi}$ | mL/g_electrode | 0.61 (theory Figs. 3 & 4) |
| | | | 0.66 (+10%=0.726) (theory Fig. 5) |

Next we describe salt adsorption in activated carbons that are charged at a fixed charging voltage of $V_{ch}$=1.2 V (with discharge at short-circuit conditions, i.e., $V_{disch}$=0 V) as function of external salt concentration. Here the geometry of a two-electrode CDI-cell is used in the experiment, where a cell voltage is applied between the two electrodes, and salt adsorption is allowed to continue until equilibrium is reached. As Fig. 3 shows, we can excellently describe both the equilibrium data for salt adsorption, SAC, and for electronic charge, $\Sigma_F$, and thus also the charge efficiency, $\Lambda$. Fig. 3b also shows the predicted $\Lambda$ when assuming zero chemical charge, which clearly overestimates the data.

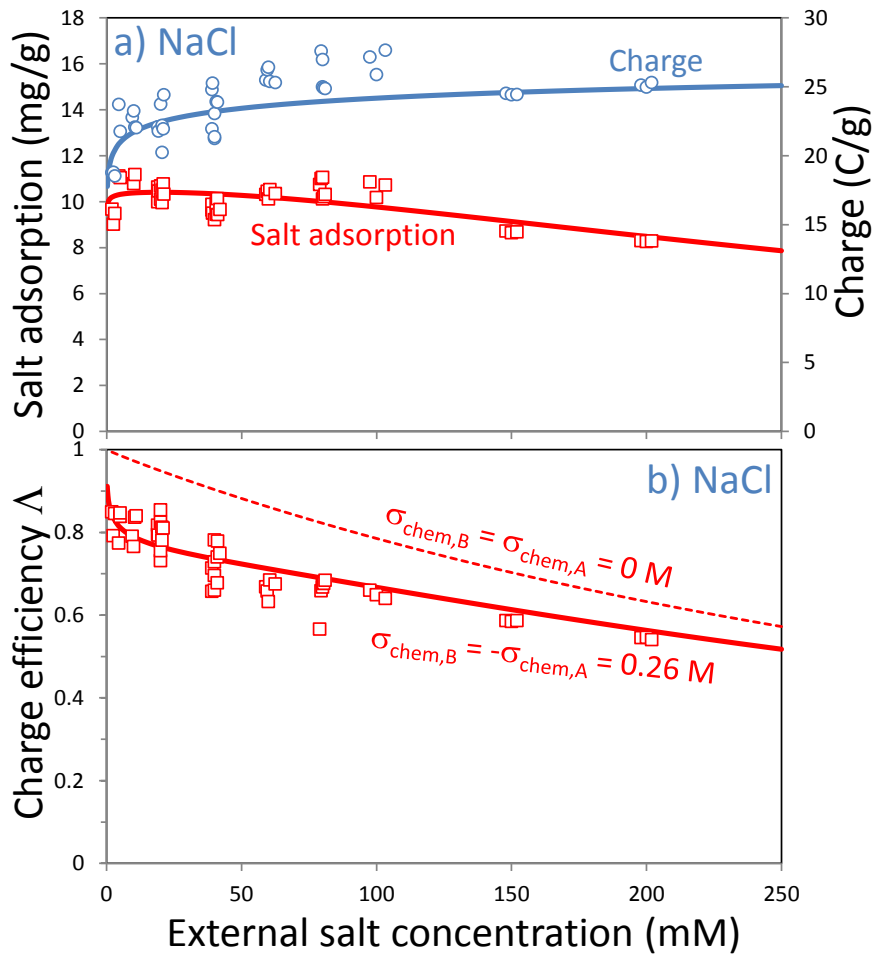

Fig. 3. a) NaCl salt adsorption and charge in CDI for charging at $V_{ch}$=1.2 V and discharge at $V_{disch}$=0 V; comparison of data (squares and circles) with amph-D model (lines). b) Data for charge efficiency compared with amph-D model, with and without chemical charge. Data from ref. [6].

Calculation results obtained by the amph-D model for SAC, $\Sigma_F$, and $\Lambda$, at varying levels of the charging voltage and for two values of the salinity, $c_\infty$, are presented in Fig. 4, and compared with data. Fig. 4 shows again how such data for the equilibrium EDL structure in microporous carbons are well described by the amph-D model. Fig. 4D shows the ion concentration in the micropores of the anode of the electrode (both data and theory based on symmetry between anode and cathode[11]) where dotted lines show the predicted concentration in each of the micro-regions, and the solid line the concentration averaged over the two regions, $<c>$, both for cations and anions ($c_\infty$=5 mM). In the cathode, the same concentrations are predicted, but with anion and cation reversed. For the concentration averaged over the pore, a similar figure is given as Fig. 3H in ref. [6].

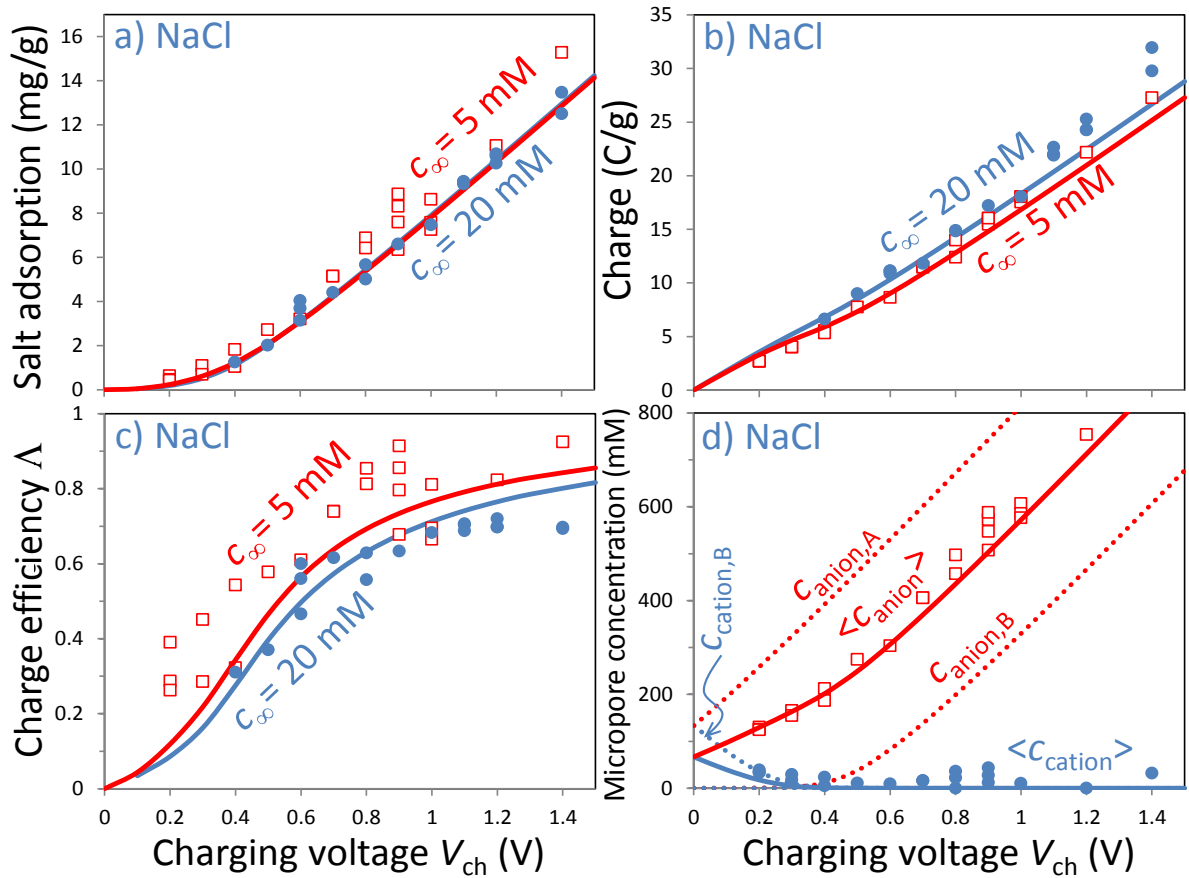

Fig. 4. a-c) NaCl adsorption, charge, and charge efficiency in CDI as function of charging voltage (discharge at $V_{disch}$=0 V); comparison of data (squares and circles) with amph-D model (lines). d) Predicted (from theory, lines) and calculated (from data, points) ion concentration in AC micropores.[6]

Next we discuss the more complicated case of asymmetric salts, such as $CaCl_2$, and mixtures of $CaCl_2$ and NaCl, as considered in ref. [12]. Data for NaCl adsorption obtained in ref. [12] can be perfectly fitted with the amph-D model, only using a slightly higher value of $v_{mi}$ than in Figs. 3 and 4, see Table 1 [not reported here], i.e., salt adsorption per gram electrode is somewhat higher than for the electrodes used in Fig. 3 and 4. Data and theory for $CaCl_2$ adsorption are given in Fig. 5. Here, we again obtain a very good fit of the model to the data, but with one reservation: in the theory the micropore volume must be increased by ~10% (compared to $v_{mi}$ required for fitting the NaCl-data in ref. [12]), see Table 1. As it were, ion adsorption for CaCl$_2$-solutions is 10% higher than theory (validated with NaCl) would predict. Fig. 5d shows a final experiment where mixtures of salts were used. Again, the amph-D model gives a very satisfactory fit to the data of equilibrium adsorption.

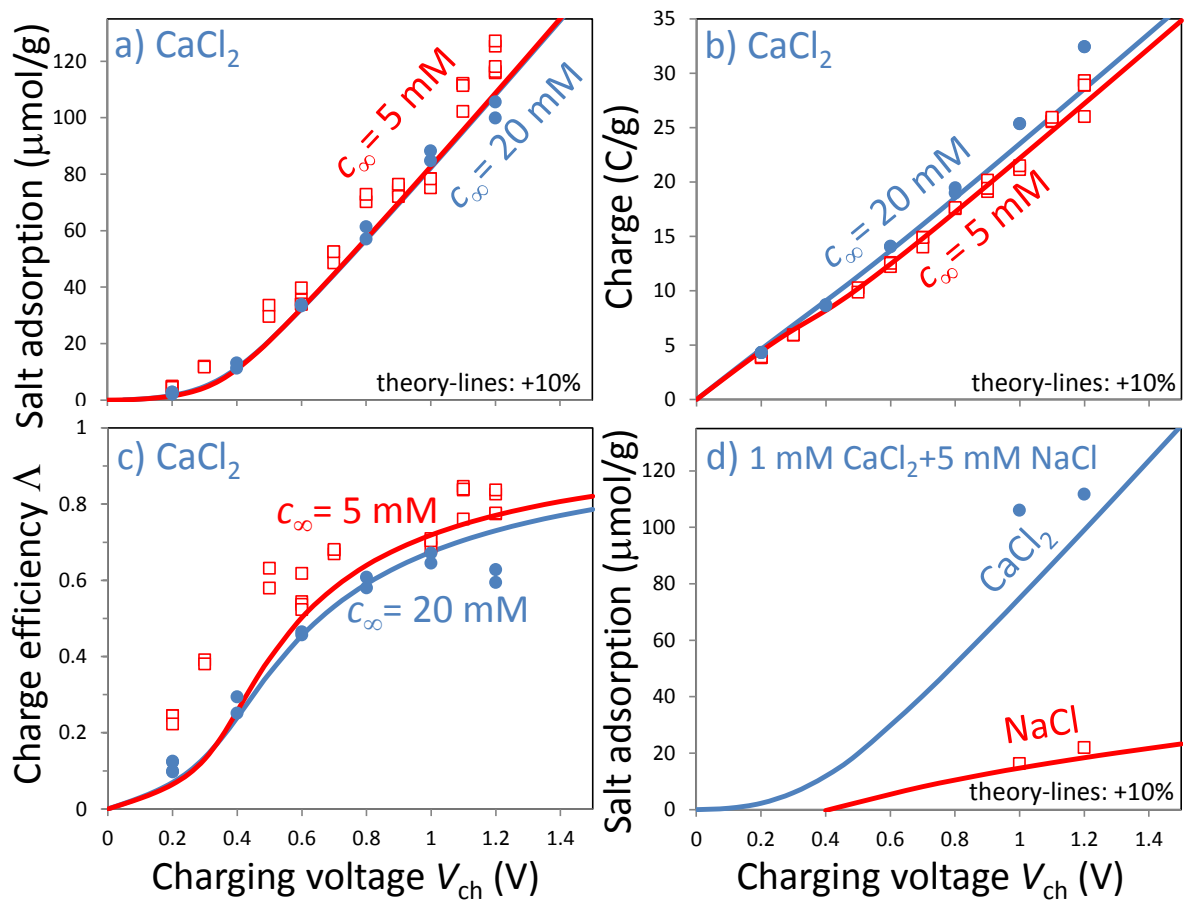

Fig. 5. a-c) CaCl$_2$ adsorption, charge, and charge efficiency in CDI as function of charging voltage (discharge at $V_{disch}$=0 V); comparison of data (squares and circles) with amph-D model (lines). d) Salt adsorption from mixture of CaCl$_2$ and NaCl. Data from ref. [12].

As a final exercise, we show in Fig. 6 calculation results where we vary the negative chemical charge, keeping the basic chemical charge at $\sigma_{chem,B}$=0.26 M. Interestingly, the calculation suggests that salt adsorption, SAC, and charge efficiency, $\Lambda$, are not at the highest value for a material symmetrically charged (see * in Fig. 6), but when the magnitude of $\sigma_{chem,A}$ is reduced, SAC and $\Lambda$ go up. Conversely, increasing $\sigma_{chem,A}$, i.e., for an electrode with more carboxylic acid groups, SAC and $\Lambda$ go down significantly, in line with results reported in ref. [24].

**Conclusions**

For ion electrosorption in activated carbon electrodes, the amphoteric Donnan model describes with a high degree of accuracy a range of data for the electrical charge and salt adsorption in aqueous solutions containing NaCl and CaCl$_2$ solutions and mixtures thereof.

**Acknowledgments**

This work was performed in the cooperation framework of Wetsus, European Centre of Excellence for Sustainable Water Technology (www.wetsus.eu). Wetsus is co-funded by the Dutch Ministry of Economic Affairs and Ministry of Infrastructure and Environment, the Province of Fryslân, and the Northern Netherlands Provinces.

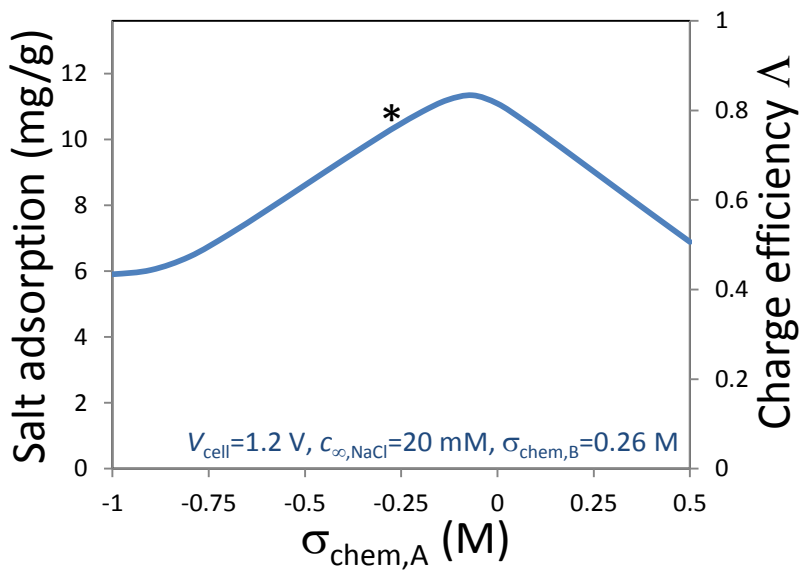

Fig. 6. Effect of chemical charge asymmetry on CDI performance. For a fixed value of the basic immobilized charge, $\sigma_{chem,B}$, the charge of negative fixed groups, $\sigma_{chem,A}$, is varied. Because charge, $\Sigma_F$, is almost invariant with $\sigma_{chem,A}$, charge efficiency, $\Lambda$, and salt adsorption, $\Gamma_{salt}$, are proportional.